\documentclass[aps,twocolumn,groupedaddress]{revtex4}
\usepackage{amssymb,amsmath}
\usepackage{graphicx,array}
\usepackage{dcolumn}
\usepackage{bm}

\begin{document}
\title{Mixing thermodynamics and photocatalytic properties of GaP--ZnS solid solutions}

\author{Joel Shenoy$^{1}$, Judy N. Hart$^{1}$, Ricardo Grau-Crespo$^{2}$, Neil L. Allan$^{3}$, and Claudio Cazorla$^{1}$}
\affiliation{$^{1}$School of Materials Science and Engineering, UNSW Sydney, NSW 2052, Australia}
\affiliation{$^{2}$Department of Chemistry, University of Reading, Reading RG6 6AD, United Kingdom}
\affiliation{$^{3}$School of Chemistry, University of Bristol, Cantock's Close, Bristol BS8 1TS, United Kingdom}

\begin{abstract}
Preparation of solid solutions represents an effective means to improve the photocatalytic properties 
of semiconductor-based materials. Nevertheless, the effects of site-occupancy disorder on the mixing  
stability and electronic properties of the resulting compounds are difficult to predict and consequently 
many experimental trials may be required before achieving enhanced photocatalytic activity. Here, 
we employ first-principles methods based on density functional theory to estimate the mixing 
free energy and the structural and electronic properties of (GaP)$_{x}$(ZnS)$_{1-x}$ solid 
solutions, a representative semiconductor-based optoelectronic material. Our method relies on a 
multi-configurational supercell approach that takes into account the configurational and vibrational
contributions to the free energy. Phase competition among the zinc-blende and wurtzite polymorphs is
also considered. We demonstrate overall excellent agreement with the available experimental data: 
(1)~zinc-blende emerges as the energetically most favorable phase, (2)~the solid solution energy band gap 
lies within the $2$--$3$~eV range for all compositions, and (3)~the energy band gap of the solid solution 
is direct for compositions $x \le 75$\%. We find that at ambient conditions most (GaP)$_{x}$(ZnS)$_{1-x}$ 
solid solutions are slightly unstable against decomposition into GaP- and ZnS-rich regions. Nevertheless, 
compositions $x \approx 25$, 50, and 75\% render robust metastable states that owing to their favorable energy 
band gaps and band levels relative to vacuum are promising hydrogen evolution photocatalysts for water 
splitting under visible light. The employed theoretical approach provides valuable insights into the 
physicochemical properties of potential solid-solution photocatalysts and offers useful guides for their 
experimental realization.   
\end{abstract}
\maketitle

\section{Introduction}
\label{sec:intro}
Storing solar energy is critical for promoting the on--demand use of renewable energy sources. 
A promising solar-energy storage approach consists of generating hydrogen fuel by photocatalytic 
water splitting under sunlight \cite{kudo08,osterloh08,tong12}. For this scheme to progress it is 
necessary to find inexpensive and efficient photocatalytic materials. Binary semiconductors, such as 
TiO$_{2}$, ZnS, and ZnO, have received great attention in this context owing to their natural 
abundance, structural simplicity, and scalable synthesis. While a direct band gap of around $2$--$3$~eV 
is most desirable for photocatalytic applications \cite{kudo08,osterloh08,tong12}, binary semiconductors 
usually present wide and/or indirect energy band gaps that limit their absorption of visible light.

Solid solutions have emerged as an effective means to improve the 
photocatalytic performance of binary semiconductors. The main idea consists of mixing isostructural 
compounds with complementary electronic properties (e.g., a wide and direct energy band gap semiconductor 
with a narrow and indirect energy band gap semiconductor) in order to breed new materials with improved 
photocatalytic performance. Examples of solid-solution photocatalysts include 
(CdS)$_{x}$(ZnS)$_{1-x}$ \cite{xing06}, (GaN)$_{x}$(ZnO)$_{1-x}$ \cite{maeda06}, 
(LaCoO$_{3}$)$_{x}$(NaTaO$_{3}$)$_{1-x}$ \cite{yi07}, and In$_{1-x}$Ni$_{x}$TaO$_{4}$ \cite{zou01}.  

\begin{figure}
\centerline{
\includegraphics[width=1.0\linewidth]{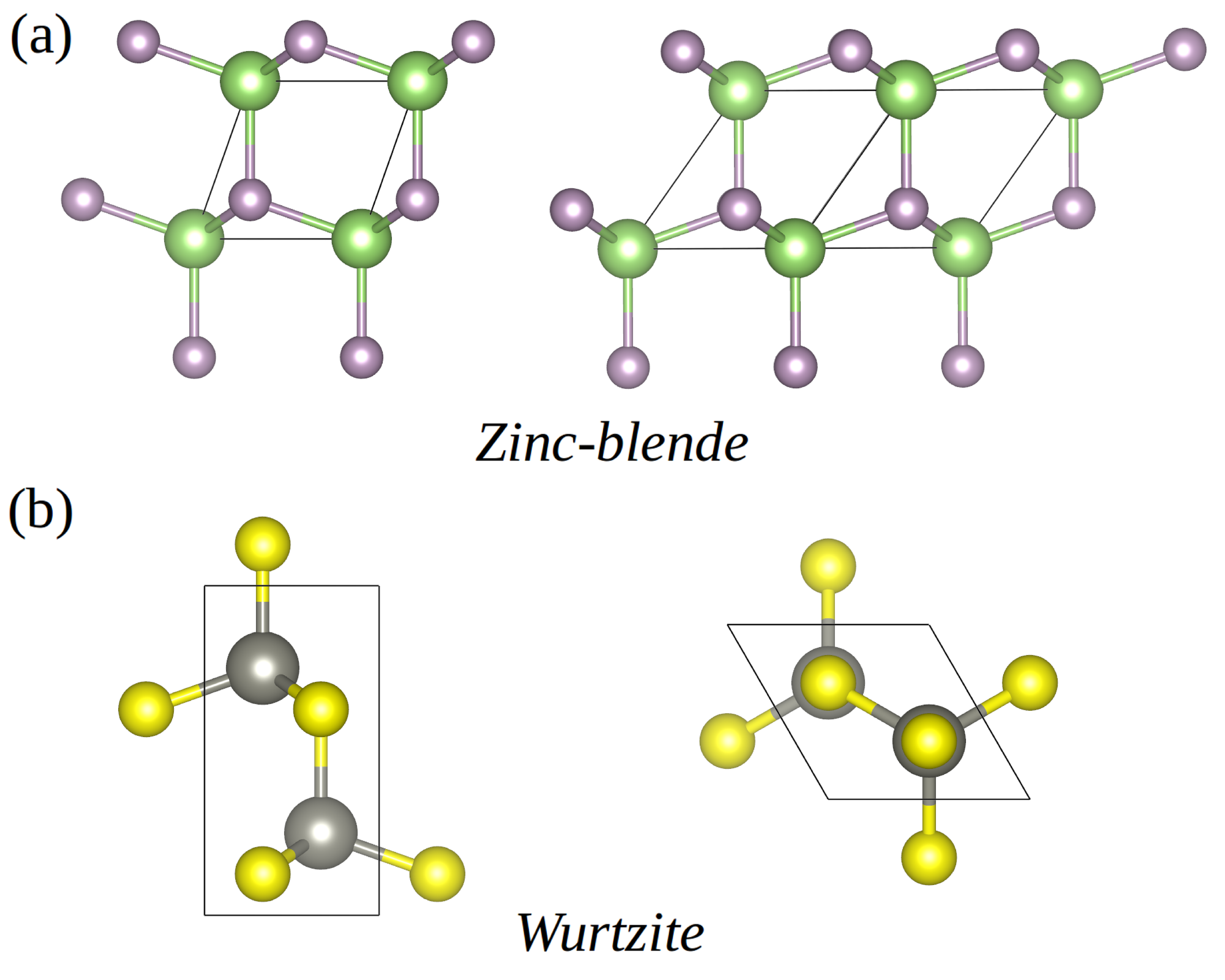}}
\caption{{\bf Polymorphism of bulk GaP and ZnS.}
         (a)~The zinc-blende structure with space group $F\overline{4}3m$ and cubic symmetry.
         (b)~The wurtzite structure with space group $P6_{3}mc$ and hexagonal symmetry.
         Ga and Zn atoms are represented with green and black spheres and P and S atoms with 
         purple and yellow spheres, respectively. In both structures the ions are four--fold 
         coordinated.}
\label{fig1}
\end{figure}

Nevertheless, the effects of site-occupancy disorder, that is, non--periodic occupation of lattice sites 
in a crystal, on the electronic properties of solid solutions are challenging to foresee; consequently, 
many experimental attempts may be required before achieving any photocatalytic enhancement. In 
addition, inhomogeneity in the synthesized materials, caused by poor thermal stability of the 
compound mixtures, may limit performance. In this regard, computer simulations are useful 
since insightful analysis of the thermodynamic, structural, and electronic properties of photocatalytic 
materials can be performed in a systematic and cost-effective manner \cite{jensen08,valentin10,hart13}. 
Nevertheless, the non-periodic occupancy of lattice sites makes modelling of solid solutions
challenging \cite{grau14,dieguez11,grau17,cazorla18}.     

Several theoretical methods have been introduced in the literature to simulate solid solutions, which can 
be classified into three main categories. The first group involves approaches in which a sort of ``average'' 
atom is defined to recover the perfect periodicity of the system, which is beneficial from a computational 
point of view. In the context of first-principles calculations, this can be achieved via the virtual-crystal 
approximation, in which the potential felt by the electrons is obtained by averaging over multiple atoms 
\cite{bellaiche00}. In the second category of methods the effects of site-occupancy disorder are straightforwardly 
reproduced by randomly occupying the lattice sites in a large periodic supercell \cite{todorov04,allan01}. 
Since the involved supercell has to be large enough to mimic a random solution, this approach may be computationally 
very intensive. A special cell generation technique was introduced by Zunger \emph{et al.} \cite{wei90} to 
construct quasi-random simulation supercells that are approachable with first-principles methods. The third 
type of solid-solution simulation method is typified by the multi-configurational supercell approach 
\cite{grau07,grau12}, in which the entire configurational space of a disordered system is first generated (for a 
finite-size supercell) and subsequently reduced to a managable set of inequivalent configurations by 
exploiting symmetry relations. Each inequivalent configuration is ascribed a Boltzmann-like occurrence 
probability that depends on its energy and configurational degeneracy, thus allowing for a full statistical
treatment of the thermodynamic and functional properties of the material. 

In this work, we present a comprehensive first-principles study of the mixing stability, and structural and 
electronic properties of (GaP)$_{x}$(ZnS)$_{1-x}$ solid solutions at temperatures $300 \le T \le 1000$~K, based 
on the multi-configurational supercell approach. GaP has an indirect energy band gap of $2.24$~eV while ZnS has 
a direct energy band gap at $\Gamma$ of $3.54$~eV, and previous computational studies have suggested that 
mixed GaP--ZnS compounds could be suitable for efficient absorption and emission of visible light \cite{hart13}. 
Indeed, (GaP)$_{x}$(ZnS)$_{1-x}$ solid 
solutions have been experimentally studied in bulk \cite{shintani73,sonomura73} and as nanowires \cite{park14}, 
showing great promise as optoelectronic materials with band gap tunability and enhanced photoluminescence
intensity. Bulk GaP and ZnS present two common polymorphs, zinc-blende 
and wurtzite (see Fig.\ref{fig1}), and in the case of (GaP)$_{x}$(ZnS)$_{1-x}$ nanowires a strong competition 
between the two phases has been reported \cite{park14}. For this reason 
we consider here both the zinc-blende and wurtzite polymorphs, which generally are observed in binary-octet semiconductors 
\cite{yeh92}. Our first-principles results show overall excellent agreement with the experimentally reported structural and electronic 
properties of GaP--ZnS solid solutions \cite{shintani73,sonomura73}. Interestingly, we predict that compositions 
$x \approx 25$, 50, and 75\% render promising photocatalyst materials for production of hydrogen fuel from water 
splitting under visible light, able to meet the requirements of (1)~a direct energy gap in the range of 
$2$--$3$~eV, (2)~a valence band level relative to vacuum lying below the water oxidation potential of
$-5.6$~eV, and (3)~a conduction band level relative to vacuum lying above the hydrogen reduction potential
of $-4.4$~eV. The solid-solution simulation approach presented in this study, therefore, is very promising 
for the design and analysis of effective photocatalytic materials, which are pressingly needed for advancing 
the field of solar-energy storage.

\section{Computational Methods}
\label{sec:methods}

\subsection{Density functional theory calculations}
\label{subsec:dft}
We use the generalised gradient approximation to density functional theory (DFT) proposed by Perdew, Burke, 
and Ernzerhof (PBE) \cite{pbe96} as implemented in the VASP software package \cite{vasp}. The projector 
augmented wave method is employed to represent the ionic cores \cite{bloch94}, considering the following 
electrons as valence states: Ga's $4s$ and $4p$; P's $3s$ and $3p$; Zn's $3d$ and $4s$; and S's $3s$ 
and $3p$. Wave functions are represented in a plane-wave basis truncated at $500$~eV. For integrations within 
the Brillouin zone (BZ) we employ Monkhorst--Pack ${\bf k}$--point grids with a density equivalent to  
$14 \times 14 \times 14$ in the unit cell. By using these parameters we obtain zero-temperature energies 
converged to within $0.5$~meV per formula unit $AB$. Geometry relaxations are performed by using a conjugate--gradient 
algorithm that allows for cell volume and shape variations; the geometry relaxations are halted after the forces 
on the atoms fall below $0.01$~eV$\cdot$\AA$^{-1}$. In order to reproduce site-occupancy disorder we adopt a 
$16$--atom simulation cell constructed by replicating $2 \times 2 \times 2$ ($2 \times 2 \times 1$) times the elemental 
$2$--atom ($4$--atom) zinc-blende (wurtzite) unit cell (see next section). 

We employ the hybrid HSE06 functional \cite{hse06} to estimate the electronic properties of equilibrium configurations 
previously generated with the PBE functional (see Supplementary Methods). This two-step approach of using PBE for 
generation of the configurational space, followed by HSE06 for electronic properties calculations, is required due to 
the high computational cost of hybrid functionals. The calculation of phonon frequencies is performed with 
the small displacement method \cite{kresse95}, in which the force-constant matrix is calculated in real-space by considering 
the proportionality between atomic displacements and forces \cite{alfe09} (see Supplementary Methods and Supplementary Fig.1). 
In order to estimate the positions relative to vacuum of the valence and conduction bands in (GaP)$_{x}$(ZnS)$_{1-x}$ solid 
solutions we employ the CRYSTAL09 code \cite{crystal09} and the hybrid B3PW functional \cite{becke93} (see Supplementary Methods).

\subsection{Multi-configurational supercell analysis}
\label{subsec:thermo}
The mixing thermodynamics of a site disordered system with a constant number of atoms and at fixed temperature
can be described as follows. For a complete set of possible system configurations ($n = 1, \ldots, N$), a 
$T$--dependent Boltzmann-like occurrence probability $p_{n}$ can be assigned to each:
\begin{equation}
p_{n} = \frac{\exp{\left( -E_{n}/k_{B}T \right)}}{Z_{\rm conf}}~,
\label{eq1}
\end{equation}
where $E_{n}$ is the corresponding energy, $k_{B}$ the Boltzmann constant, and $Z$ the partition function defined 
as:
\begin{equation}
Z_{\rm conf} = \sum_{n=1}^{N}\exp{\left( -E_{n}/k_{B}T \right)}~.
\label{eq2}
\end{equation} 
Accordingly, the Helmholtz free energy of the system $F$ per formula unit (f.u.) can be estimated as:
\begin{equation}
F_{\rm conf} = -\frac{1}{N_{\rm f.u.}}k_{B}T \ln Z_{\rm conf}~,
\label{eq3}
\end{equation}
where $N_{\rm f.u.}$ is the number of formula units in the simulation cell (equal to the total
number of atoms divided by two). The average value of the energy (or of any other well-defined 
quantity $A$ for each configuration $n$) in the corresponding configurational 
space adopts the form:
\begin{equation}
\langle E \rangle = \sum_{n=1}^{N} p_{n}E_{n} \qquad \left( \langle A \rangle = \sum_{n=1}^{N} p_{n}A_{n}  \right)~.
\label{eq4}
\end{equation}

In practice the configurational space of a chemically disordered solid may comprise a huge number of configurations ($N \gg 10^{3}$), 
hence normally it cannot be described with first-principles methods. An effective way to overcome such a limitation is to reduce 
the total number of possible configurations to $M \ll N$ by exploiting the symmetry properties 
of the parent crystal phase \cite{grau07}. In the generated reduced configurational space each inequivalent configuration 
$m$ has an associated degeneracy $\Omega_{m}$, equal to the number of symmetrically equivalent configurations with same energy 
$E_{m}$, which all together fulfil the relation $N = \sum_{m=1}^{M} \Omega_{m}$. The average value of the energy (or of any 
other well-defined quantity $A$ for each configuration $n$) in the reduced configurational space then is 
expressed as:
\begin{equation}
\langle E \rangle = \sum_{m=1}^{M} \tilde{p}_{m}E_{m} \qquad \left( \langle A \rangle = \sum_{m=1}^{M} \tilde{p}_{m}A_{m}  \right)~,
\label{eq5}
\end{equation}
where:
\begin{equation}
\tilde{p}_{m} = \frac{1}{Z_{\rm conf}}\Omega_{m}\exp{\left( -E_{m}/k_{B}T \right)}~.
\label{eq6}
\end{equation}

The Helmholtz free-energy function in Eq.(\ref{eq3}) accounts for configurational effects but neglects 
contributions from $T$--induced lattice excitations, which \emph{a priori} may be important in solid 
solutions \cite{walle02}. To include vibrational contributions in our free-energy calculations, we 
adopt the expression:
\begin{equation}
F = F_{\rm conf} + F_{\rm vib}~,
\label{eq7}
\end{equation}
where \cite{cazorla17,cazorla13,cazorla17b}:
\begin{eqnarray}
F_{\rm vib} (T) & = & \frac{1}{N_{\rm f.u.} N_{q}}~k_{B} T \times \nonumber \\ 
& & \sum_{{\bf q}s}\ln\left[ 2\sinh \left( \frac{\hbar \omega_{{\bf q}s}}{2k_{\rm B}T} \right) \right]~.
\label{eq8}
\end{eqnarray}
In Eq.(\ref{eq8}), $N_{q}$ represents the total number of wave vectors used for integration within the BZ, $\omega_{{\bf q}s}$ 
the vibrational eigenfrequencies of the system, and the summation runs over all wave vectors ${\bf q}$ and phonon 
branches $s$. Due to the huge computational expense associated with first-principles estimation of phonon excitations 
for a large number of configurations, we calculate $F_{\rm vib}$ just for the structure rendering the highest $\tilde{p}_{m}$ 
probability at room temperature (see Eq.\ref{eq6}). We justify this choice in detail later when discussing
our results; nevertheless, we estimate that the $F_{\rm vib}$ errors resulting from this procedure are
 within $10$~meV per formula unit (see Supplementary Methods).  

To assess the mixing stability of bulk GaP--ZnS solid solutions, and since we consider zero-pressure conditions
in this work, we estimate the corresponding mixing free-energy $\Delta F$ as a function of temperature and 
composition:
\begin{equation}
\Delta F (x, T) = F^{\rm ss}(x,T) - xF^{\rm GaP}(T) - (1-x)F^{\rm ZnS}(T)~, 
\label{eq9}
\end{equation}  
where the superscripts indicate the system for which the Helmholtz free energy is calculated (``ss'' stands 
for solid solution). A disordered system is thermodynamically stable against decomposition into GaP- and ZnS-rich
regions if $\Delta F (x, T) < 0$ (although it may be unstable with respect to fluctuations in composition if 
$\frac{\partial^2 \Delta F (x, T)}{\partial x^2} < 0$); otherwise, the system may be thermodynamically metastable 
or unstable (depending on whether $\frac{\partial^2 \Delta F (x, T)}{\partial x^2}$ is positive or negative).
In this work, we perform the multi-configurational supercell calculations and accompanying statistical analysis with the 
SOD software \cite{grau07}. We use a $16$--atom supercell that allows explicit simulation of 
(GaP)$_{x}$(ZnS)$_{1-x}$ solid solutions at $9$ different compositions (namely, $x_{k} = \frac{k}{8}$ with $k = 0, \ldots, 8$); 
results at other intermediate compositions are obtained via smooth spline interpolations. The resulting finite-size  
$\Delta F$ bias is estimated to be of the order of $10$~meV per formula unit (see tests performed for a larger 
supercell containing $24$ atoms explained in the Supplementary Methods).  

\begin{figure*}
\centerline{
\includegraphics[width=1.0\linewidth]{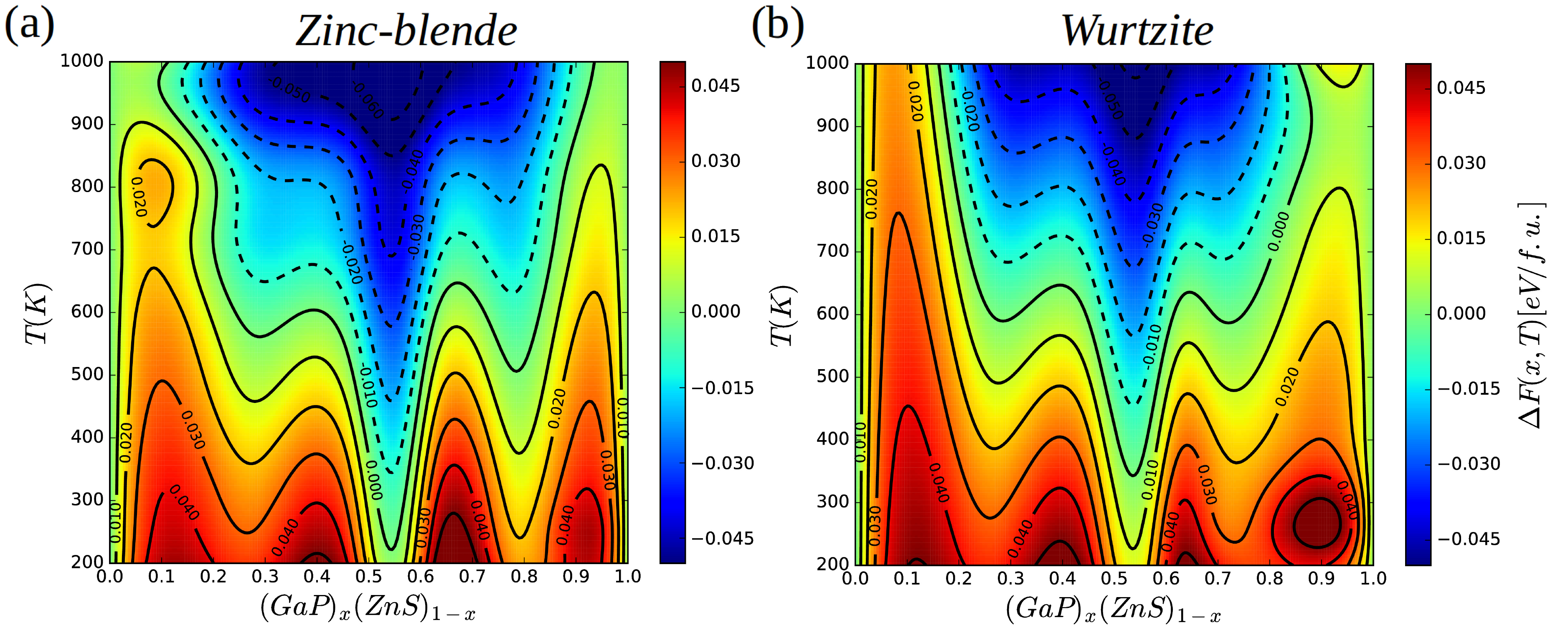}}
\caption{{\bf Mixing free-energy of (GaP)$_{x}$(ZnS)$_{1-x}$ solid solutions as a function of temperature and composition
	 (color online).} (a)~Zinc-blende and (b)~wurtzite structures. Solid lines represent isovalue $\Delta F$ contours 
	 expressed in units of eV per formula unit. Typical $\Delta F$ errors are estimated to be of the order of $10$~meV 
	 per formula unit.}
\label{fig2}
\end{figure*}

Finally, we note that in the limit of very high temperatures the configurational entropy $S_{\rm conf}$ equals 
$\frac{1}{N_{\rm f.u.}}k_{B}\ln{N}$. Due to the finite size of the employed simulation cell, such a configurational 
entropy limit generally underestimates (in absolute value) the quantity: 
\begin{equation}
S_{\rm conf}^{\infty} = -2k_{B}\left( x\ln{x} + \left( 1-x \right)\ln{\left(1-x\right)}  \right)~,
\label{stirling}
\end{equation}
which is exact in the thermodynamic limit ($N \to \infty$ at constant composition $x$) and $T \to \infty$
(the factor ``2'' in the formula above appears due to occupancy disorder in both the anion and cation sublattices). 
Aimed at correcting for such unavoidable finite-size bias, and as has been done in previous work \cite{grau12,becker00}, 
we apply the following shift to the configurational free energy:
\begin{equation}
F_{\rm conf}^{\rm corr} = F_{\rm conf} - T(S_{\rm conf}^{\infty} - \frac{1}{N_{\rm f.u.}} k_{B}\ln{N})~. 
\label{corr}
\end{equation}   
By doing this, the correct expression of the configurational free energy is consistently recovered in the $T \to \infty$ 
limit. We note that, while this correction is quantitatively significant, the main conclusions presented in the following 
sections are not qualitatively affected by it (see Fig.\ref{fig2} and Supplementary Fig.2). 

\begin{figure*}
\centerline{
\includegraphics[width=1.0\linewidth]{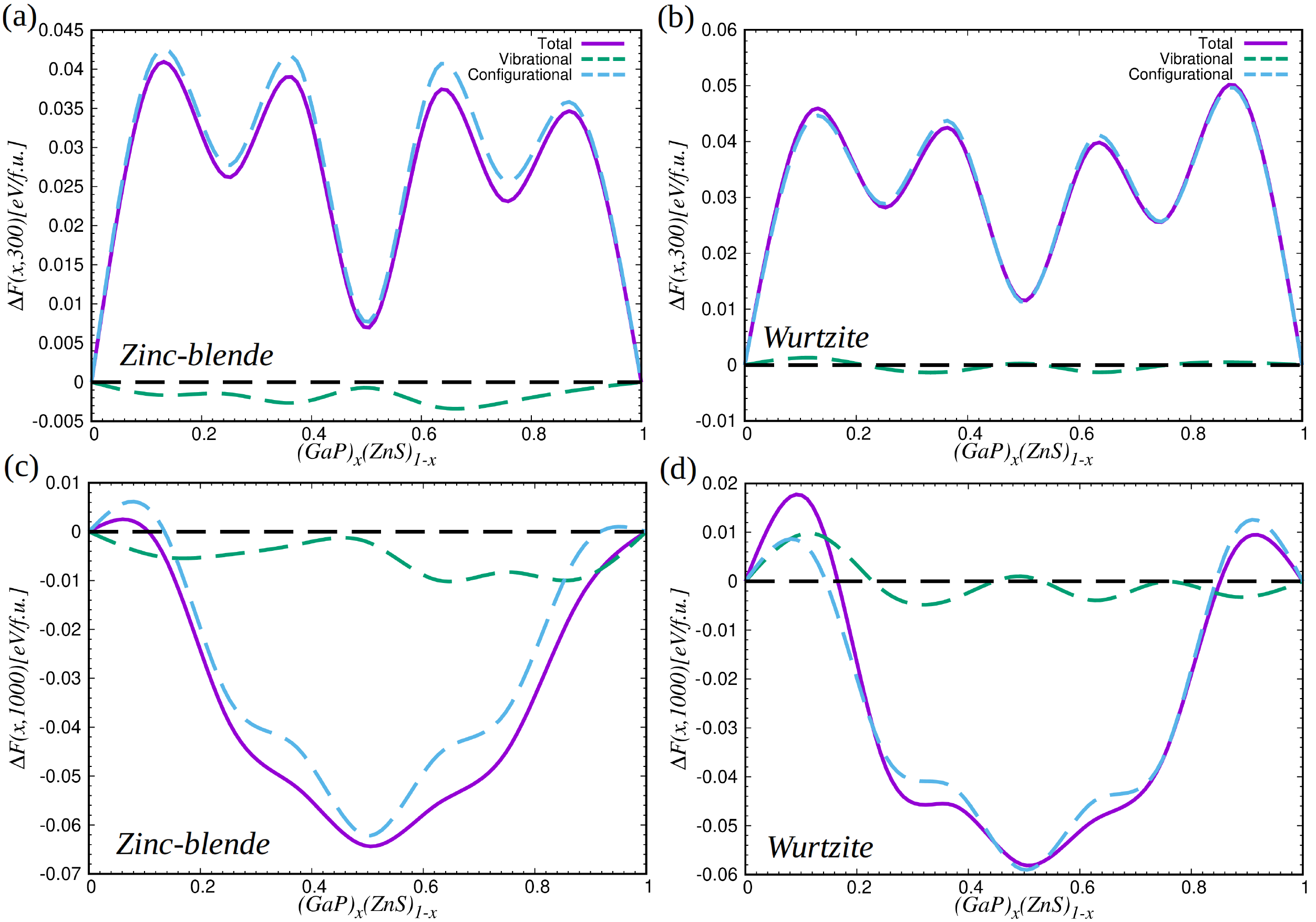}}
\caption{{\bf Mixing free-energy of (GaP)$_{x}$(ZnS)$_{1-x}$ solid solutions at 300 and 1000~K.}
        (a)-(c)~Zinc-blende and (b)-(d)~wurtzite structures. Vibrational and configurational (i.e., 
        total minus vibrational) contributions to the total mixing free--energy of the system are 
        expressed as a function of composition. Typical $\Delta F$ errors are estimated to be of the 
        order of $10$~meV per formula unit.}
\label{fig3}
\end{figure*}

\begin{figure}[t!]
\centerline{
\includegraphics[width=1.0\linewidth]{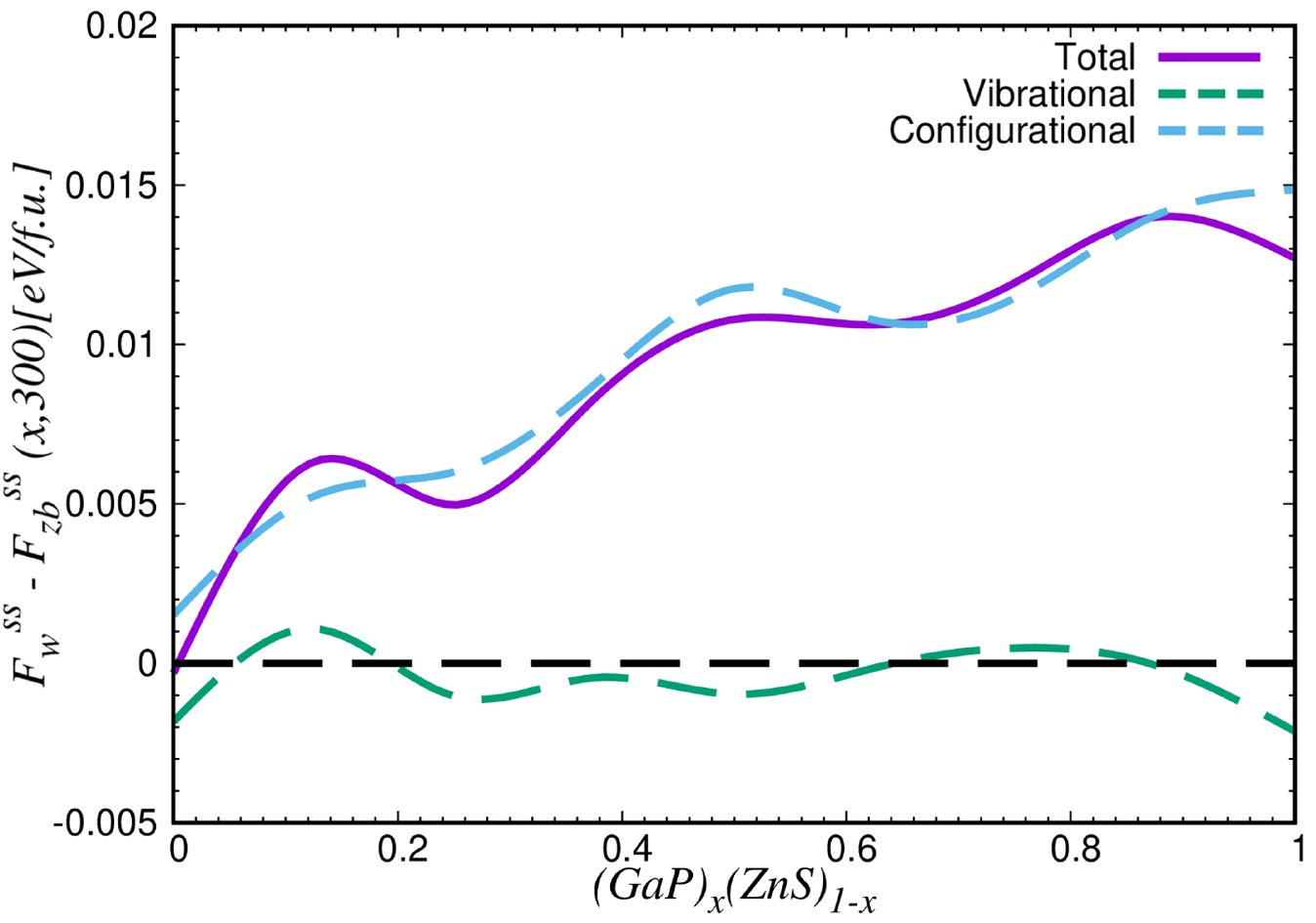}}
\caption{{\bf Free-energy difference between the two polymorphs for (GaP)$_{x}$(ZnS)$_{1-x}$ solid solutions 
	 at $T = 300$~K.} Zinc-blende and wurtzite phases are represented as ``zb'' and ``w'', respectively.
          Configurational and vibrational contributions to the total free--energy difference are indicated.
	  Note that for the end-members $F_{\rm conf}$ reduces to the static energy (see Eqs.\ref{eq2}--\ref{eq3}).} 
\label{fig4}
\end{figure}

\begin{figure}[t!]
\centerline{
\includegraphics[width=1.0\linewidth]{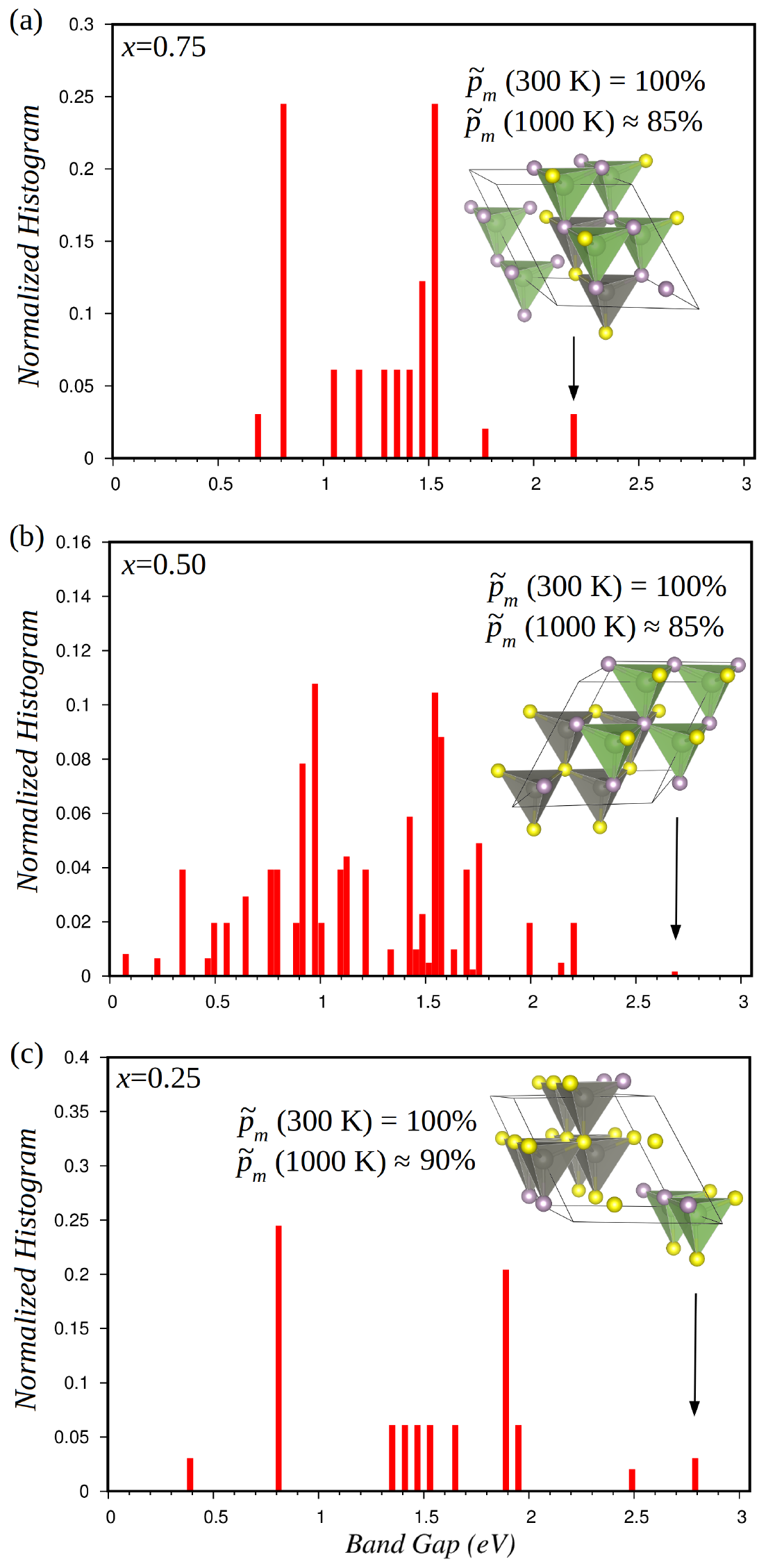}}
\caption{{\bf Normalized $E_{\rm gap}$ histogram for zinc-blende (GaP)$_{x}$(ZnS)$_{1-x}$ 
         solid solutions.} (a)~$x = 75$, (b)~$50$, and (c)~$25$\%~compositions. The histogram is constructed 
	 by dividing the number of structures within each specific band gap energy interval by the total 
	 number of structures. The inequivalent configuration that is most probable in each case is sketched 
	 along with its occurrence probability and energy band gap (black arrow). Ga and Zn atoms are represented 
	 with green and black spheres and P and S atoms with purple and yellow spheres, respectively. Energy 
	 band gaps are estimated with the HSE06 hybrid functional \cite{hse06}.}
\label{fig5}
\end{figure}

\section{Results and Discussion}
\label{sec:results}

\subsection{Mixing thermodynamics}
\label{subsec:tstability}
Figure~\ref{fig2} shows the mixing free-energy of GaP--ZnS solid solutions calculated as a function of 
structure, composition, and temperature at zero-pressure conditions (thermal expansion effects have been
neglected). We have selected the temperature interval $200 \le T \le 1000$~K because typical synthesis 
temperatures of semiconductor-based solid solutions are of the order of $10^{2}$--$10^{3}$~K. 

Below ambient conditions, we find that all zinc-blende bulk systems are thermodynamically unstable against 
decomposition into the end-members GaP and ZnS since $\Delta F > 0$, although some compositions may be 
kinetically stable where $\frac{\partial^2 \Delta F (x, T)}{\partial x^2}$ is locally positive (Fig.\ref{fig2}a). 
However, at temperatures moderately above $T = 300$~K some disordered crystals with compositions close to $x= 50$\% 
become thermodynamically stable against decomposition into end-members GaP and ZnS, since $\Delta F < 0$. The 
composition range of this stability increases with increasing temperature, reaching $15 \le x \le 85$\% at 1000~K.
Therefore, there exists the possibility of generating metastable solid solutions in that range of compositions 
at ambient temperature via quenching \cite{jones73}. The most favorable compositions for the realization of such 
metastable states are for $x$ close to $25$, $50$, and $75$\% since at a fixed temperature they render local 
minima in compositional space (hence $\frac{\partial^2 \Delta F (x, T)}{\partial x^2}$ is negative, so 
decomposition into GaP- and ZnS-rich regions can be kinetically hindered since small fluctuations in composition 
result in an increase in free energy). On the contrary, compositions near $x \approx 15$, $35$, $65$ and $85$\% 
appear to be thermodynamically most unstable (since they render local maxima in compositional space, where 
$\frac{\partial^2 \Delta F (x, T)}{\partial x^2}$ is negative, and hence small fluctuations in composition 
result in a decrease in free energy). 

The mixing properties of GaP--ZnS solid solutions in the wurtzite structure are very similar to those just 
described for the zinc-blende polymorph (Fig.\ref{fig2}b). The same most favorable and most disadvantageous 
compositions for the synthesis of metastable systems are found, and at low temperatures the two corresponding 
$\Delta F$ maps are practically identical. However, as temperature is increased the mixing free-energy of the 
zinc-blende phase becomes slightly more favorable than that of the wurtzite polymorph. For instance, at 
$T = 1000$~K and $x = 50$\% we estimate $\Delta F = -0.058$~eV/f.u. in the wurtzite phase and $-0.065$~eV/f.u. 
in the zinc-blende phase. 

To understand the origins of the mixing stability differences between the zinc-blende and wurtzite polymorphs,
 we plot in Fig.\ref{fig3} the corresponding $\Delta F$ curves calculated at $T = 300$ and $1000$~K,  
split into different contributions (i.e., total, vibrational, and configurational). At room temperature (see 
Figs.~\ref{fig3}a-b) the vibrational contribution to the mixing free-energy is practically negligible in both 
phases (see green lines therein), hence configurational effects are the dominant cause of the $\Delta F$ variations 
observed across the composition series. However, at high temperatures the vibrational mixing free-energy 
has a stabilizing effect in both phases, especially at compositions $x > 50$\%, and is largest in absolute value in the 
zinc-blende polymorph (see Figs.~\ref{fig3}c-d). For instance, at $T = 1000$~K and $x \approx 60$\% $\Delta F_{\rm vib}$ 
amounts to $-10$~meV/f.u. in the zinc-blende phase and to $-4$~meV/f.u. in the wurtzite phase. The reason for 
such a vibrational mixing free-energy difference is that the high-frequency phonons, which are dominant at high 
temperatures, present lower energies in the zinc-blende phase (see Supplementary Fig.1). Meanwhile, configurational 
effects (blue lines in the figures) also tend to favor slightly the cubic polymorph over the hexagonal (e.g., 
at $T = 300$~K and $x \approx 60$\% $\Delta F_{\rm conf} = 30$~meV/f.u. in the zinc-blende phase and $40$~meV/f.u. 
in the wurtzite phase). 

Figure~\ref{fig4} shows the free-energy difference between GaP--ZnS solid solutions in the wurtzite and zinc-blende 
phases expressed as a function of composition at $T = 300$~K. It is appreciated that the zinc-blende polymorph
is energetically more favorable than the wurtzite phase at all compositions. Equivalent results are obtained
also at higher temperatures (see Supplementary Fig.3). The main reason for the preference of the zinc-blende 
phase is that the configurational free energy is larger in absolute value than in the cubic polymorph. This is 
clearly appreciated in Fig.\ref{fig4}, where we show that the vibrational free-energy differences between the two 
polymorphs are practically negligible as compared to the configurational counterparts (see blue and green lines therein). 
Hence our results indicate that the zinc-blende phase is more likely to be synthesized in practice than
the wurtzite polymorph, which is consistent with the experimental observations \cite{shintani73,sonomura73}. In the 
remainder of the article, we will focus on zinc-blende solid solutions at compositions $x = 25$, $50$, and $75$\% 
since these render the best thermodynamic stabilities. 

\begin{figure*}
\centerline{
\includegraphics[width=1.0\linewidth]{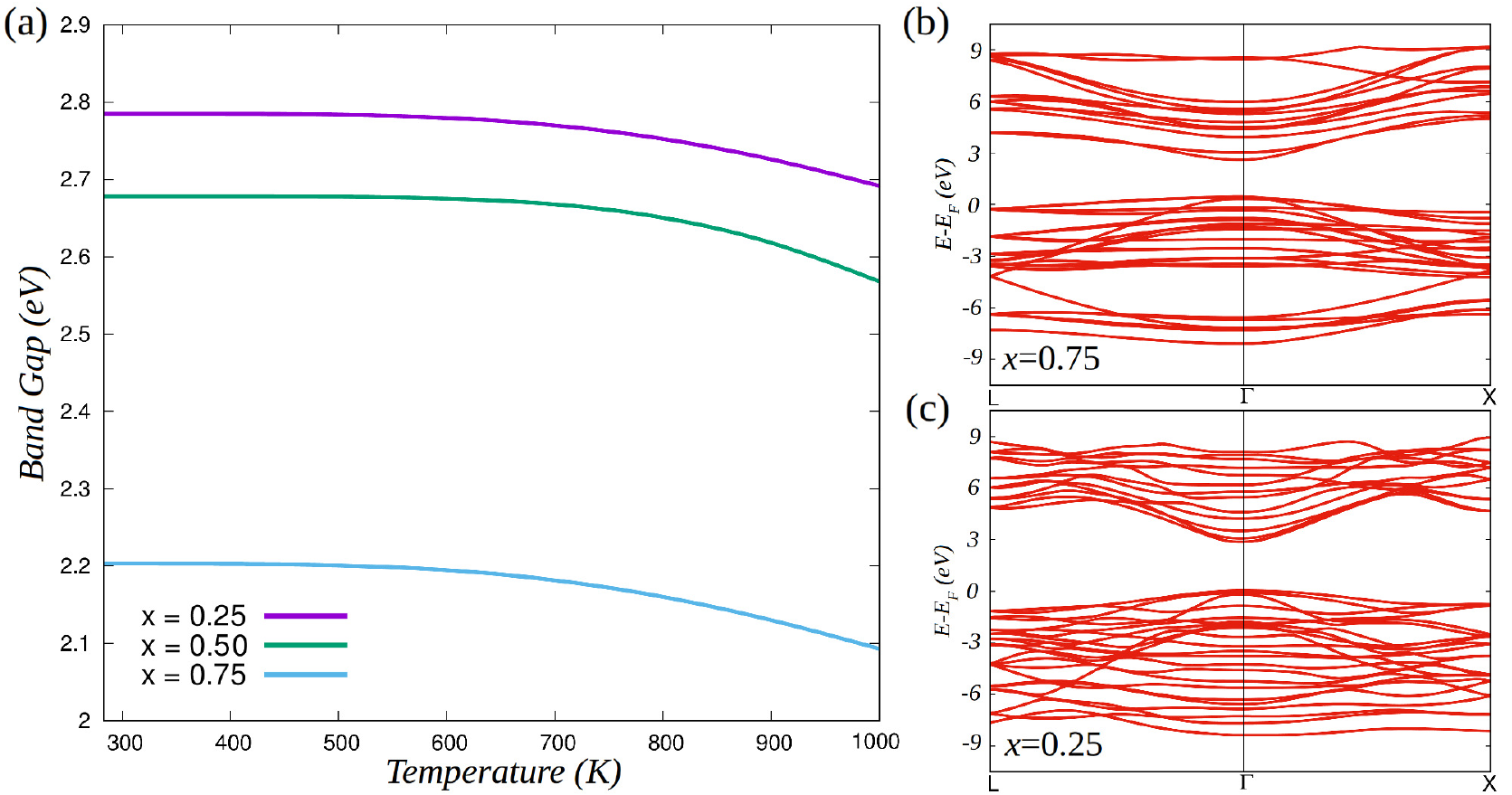}}
\caption{{\bf Electronic band structure of zinc-blende (GaP)$_{x}$(ZnS)$_{1-x}$ solid solutions.} 
         (a)~Averaged energy band gap as a function of temperature and composition calculated with 
	 the multi-configurational supercell method. (b)~Electronic band structure calculated
         for the most probable configuration at $x = 75$\%, which renders a direct energy band gap 
	 at $\Gamma$. (c)~Idem at $x = 25$\%. Energy band gaps are estimated with the exchange-correlation 
	 HSE06 hybrid functional \cite{hse06}.}
\label{fig6}
\end{figure*}

\begin{figure}
\centerline{
\includegraphics[width=1.0\linewidth]{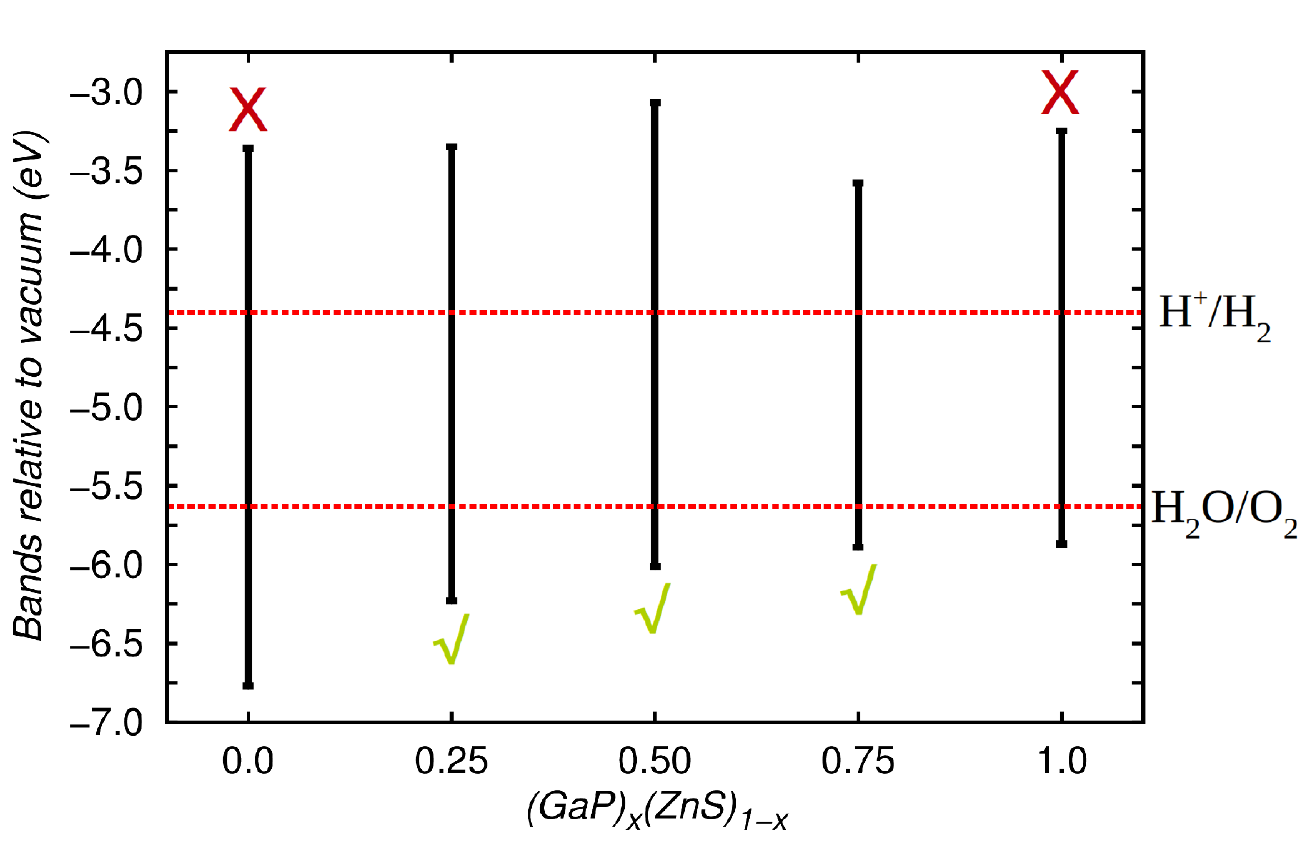}}
\caption{{\bf Energy bands relative to the vacuum level in zinc-blende (GaP)$_{x}$(ZnS)$_{1-x}$ solid solutions.}
         $\Gamma$ energy band gaps are direct for compositions $x = 0$, 25, 50, and 75\%. All energy
         band gaps lie within the $2$--$3$~eV interval except for ZnS. The redox potentials relative to
         the vacuum level for the hydrogen reduction and water oxidation reactions are indicated with
         dashed lines. Energy band gaps and levels are estimated with the B3PW hybrid functional \cite{becke93}
         and CRYSTAL09 code \cite{crystal09}.}
\label{fig7}
\end{figure}

\subsection{Electronic properties}
\label{subsec:optoelecat}

\subsubsection{Energy band gap}
\label{subsec:egap}
In Fig.\ref{fig5}, we show the normalized energy band gap histograms calculated with the HSE06 hybrid 
functional \cite{hse06} for zinc-blende solid solutions (Supplementary Methods). Specifically, $E_{\rm gap}^{\rm HSE06}$ 
is computed for each structure in the corresponding $x$ configurational ensemble and a normalized 
histogram is constructed subsequently by considering small energy increments (i.e., the number of 
structures within each specific band gap energy range is divided by the total number of structures 
--the occurrence probabilities $\tilde{p}_{m}$ are not considered here--). 
A large dispersion of energy band gap values spanning over the interval $0.50 \lesssim E_{\rm gap}^{\rm HSE06} 
\lesssim 2.75$~eV is observed. In the $x = 50$\% case (Figs.~\ref{fig5}b), we find particularly small $E_{\rm gap}^{\rm HSE06}$ 
values, that is, below $0.50$~eV, and the normalized histograms are quasi-continuous owing to the higher 
degree of site-occupancy disorder. It is worth noting that all the calculated energy band gaps are smaller than that 
of pure ZnS (i.e., $E_{\rm gap}^{\rm exp} = 3.5$~eV \cite{heyd05} and $E_{\rm gap}^{\rm HSE06} = 3.1$~eV) and 
lie within the range of visible or infrared light. Analogous $E_{\rm gap}^{\rm HSE06}$ results are obtained also
for the wurtzite polymorph (see Supplementary Fig.4).  
 
The large dispersions shown in Fig.\ref{fig5} indicate a strong dependence of the energy band gap, and in general of the 
electronic band structures of (GaP)$_{x}$(ZnS)$_{1-x}$ solid solutions, on the local atomic environment. 
Similar behaviour is likely to occur in analogous binary octet semiconductors.
For the zinc-blende (GaP)$_{0.5}$(ZnS)$_{0.5}$ system, we have performed a detailed analysis of the 
density of electronic states considering different atomic arrangements to better understand the origins of the 
$E_{\rm gap}$ variations. A clear electronic--structural correlation emerges from our calculations: the larger (smaller) 
the number of Zn--S bonds, or Ga--P bonds, the larger (smaller) the energy band gap. In turn, the electrostatic
potential produced by local environments can be directly related to the number of Zn--S, or Ga--P, bonds. 
Specifically, when the number of Zn--S bonds, or Ga--P bonds, is minimized (leading to the smallest $E_{\rm gap}$) the 
electrostatic potential at the S and P atoms forming the top of the valence band (see Supplementary Fig.5) is lowest (i.e., 
their atomic environment is most negatively charged), 
hence the energy of the valence band is high. Likewise, the electrostatic potential at the Ga atoms forming the bottom 
of the conduction band (see Supplementary Fig.5) is highest when the number of Ga--P bonds, or Zn--S bonds, is minimized, 
which induces a lowering in the energy of the conduction band. 

Remarkably, however, when the average value of $E_{\rm gap}$ 
is calculated with the multi-configurational supercell approach (see Eqs.\ref{eq5}--\ref{eq6} in 
Sec.~\ref{subsec:thermo}), the band gap dispersions shown in Fig.\ref{fig5} play just a marginal role. As we explain 
next, this is due to the fact that one particular configuration is much more likely to occur than the rest 
(see the occurrence probabilities indicated in Fig.\ref{fig5} and the full $\tilde{p}_{m}$ spectra shown in 
Supplementary Fig.6).   

Figure~\ref{fig6} shows the $T$-dependent average value of the energy band gap calculated with the multi-configurational 
supercell method (see Eqs.(\ref{eq5})--(\ref{eq6})) 
for GaP--ZnS solid solutions at $x = 25$, $50$, and $75$\%. $E_{\rm gap}$ presents an almost constant value throughout the 
whole temperature interval $300 \le T \le 1000$~K at any composition, and increases from $2.1$ to $2.8$~eV as $x$ decreases 
from $75$ to $25$\%. As could have been foreseen, the energy band gap of the mixed compound is larger when the content of ZnS 
is larger (recall that $E_{\rm gap}^{\rm exp} = 2.3$ and $3.5$~eV in pure GaP and ZnS \cite{hart13,heyd05}, respectively). The 
reason for the steady behaviour of $E_{\rm gap}$ as a function of temperature is that in all the analyzed cases the spectra 
of occurrence probabilities $\tilde{p}_{m}$ (see Eq.~\ref{eq6}) very much favour the configuration with lowest energy (and low 
configurational degeneracy as shown by the histograms in Fig.\ref{fig5}) over the others (Supplementary Fig.6). The lowest-energy 
configurations determined at each $x$ are sketched in Fig.\ref{fig5}, along with their energy band gaps (see black arrows therein). 
As can be observed, the lowest-energy configurations render structures in which the number of Zn--S, or Ga--P, bonds are maximized, 
which as we have explained before correlates directly to largest energy band gaps. 

The small $E_{\rm gap}$ variations observed in Fig.\ref{fig6}a at $T > 600$~K 
stem from the increasingly more important role that highly degenerate configurations (i.e., with large $\Omega_{m}$'s --see 
Sec.~\ref{subsec:thermo}--) start to play at high temperatures (Supplementary Fig.6). Analogous energy band gap results 
are obtained also for the wurtzite phase (see Supplementary Fig.7). It is worth noting that the highly peaked nature of the 
$\tilde{p}_{m}$ distributions calculated in GaP--ZnS solid solutions (see Supplementary Fig.6) comes to justify the strategy 
that we have followed for estimating vibrational contributions to the mixing free-energy (see Sec.~\ref{subsec:thermo}).  

The energy band gap results shown in Fig.\ref{fig6} are in very good agreement with the optical measurements performed
by Shintani and Sonomura more than $40$ years ago \cite{shintani73,sonomura73}. In particular, the room-temperature experimental 
results $E_{\rm gap}^{\rm exp} = 2.4$--$2.6$, $2.4$--$2.5$, and $2.2$--$2.4$~eV obtained at $x = 25$, $50$, and $75$\%, respectively, 
compare very well to our theoretical values $E_{\rm gap}^{\rm HSE06} = 2.80$, $2.65$, and $2.20$~eV. Moreover, in our simulations 
we find that all compositions $x \le 75$\% render direct energy band gaps at $\Gamma$ (see Figs.\ref{fig6}b,c, obtained for the 
most probable configurations), which also is consistent with the reported experimental obervations \cite{shintani73,sonomura73}. 
Thus, we confirm that zinc-blende (GaP)$_{x}$(ZnS)$_{1-x}$ solid solutions display promising visible-light absorption features, 
namely, direct energy band gaps lying in the range $2$--$3$~eV \cite{hart13} (likewise, the wurtzite polymorph possesses also 
promising electronic properties, see Supplementary Fig.7). Overall, our theoretical findings demonstrate that the present 
computational approach, which is based on the multi-configurational supercell method and density functional theory calculations, 
is able to reproduce closely the electronic band structure features of complex materials with site-occupancy disorder.

\subsubsection{Energy band levels relative to vacuum}
\label{subsec:ebandlevel}
Promising photocatalytic materials should present not only suitable energy band gaps but also adequate energy
band levels relative to vacuum \cite{moses11}. Here, we are interested in assessing the potential of 
(GaP)$_{x}$(ZnS)$_{1-x}$ solid solutions for production of hydrogen fuel from water splitting under visible light. 
To this end, we calculated the position of the top of the valence band (VB) and bottom of the conduction band (CB) 
relative to the vacuum energy using the CRYSTAL09 code \cite{crystal09} with the methods described in Sec.~\ref{subsec:dft} 
and Supplementary Methods at compositions $x = 0$, $25$, $50$, $75$, and $100$\%. In particular, we calculated the difference 
between the energy of the $1s$ orbitals in bulk GaP and bulk ZnS and in the centre of ZnS and GaP slabs, $\Delta E_{1s}$. 
A non-polar $(110)$--oriented slab with a total thickness equivalent to $20$ atomic layers was used for these calculations 
(corresponding to a thickness of $37$~\AA~ for ZnS and of $38$~\AA~ for GaP; such a length provides sufficiently well-converged 
results). The $(110)$ surface was selected because this has been reported to be the most stable for zinc-blende ZnS \cite{hamad02,wright98} 
and GaP \cite{hayashi82}. The cell length in the direction perpendicular to the slab surface was set to $500$~\AA, giving a 
large enough vacuum gap to prevent interactions between periodic images of the slab. The slab was fully relaxed in all cases.
The energy difference $\Delta E_{1s}$ was then added to the VB and CB energies in order to obtain the energies relative to vacuum. 
For the solid solutions, the shift applied to the band energies was a weighted average of those for pure ZnS and pure GaP. This 
method has been previously applied to GaN--ZnO solid solutions \cite{valentin10}. 

Figure~\ref{fig7} shows our hybrid B3PW \cite{becke93} results for the energy band levels relative to vacuum for 
different compositions. We reiterate that an ideal hydrogen photocatalyst material should present (1)~a direct energy band 
gap lying in the range $2$--$3$~eV, (2)~a VB level relative to vacuum lying below the water oxidation potential of $-5.6$~eV, 
and (3)~a CB level relative to vacuum lying above the hydrogen reduction potential of $-4.4$~eV. First, we note that our band 
alignment results obtained for bulk GaP and ZnS are in good agreement with previous first-principles calculations reported 
by other authors \cite{vandewalle03}. Pure GaP and ZnS are not suitable for photocatalytic water splitting 
since they do not fulfill condition (1) above. In contrast, all three solid solutions $x = 25$, $50$, and $75$\% fulfill 
requirements (1)--(3) and hence are promising hydrogen photocatalyst materials for water splitting. For instance, in the 
$x = 75$\% ($x = 25$\%) case the VB level lies $0.3$ ($0.6$)~eV below $-5.6$~eV and the CB level is $0.8$ ($1.0$)~eV 
above $-4.4$~eV. We note that metastable ZnS-rich solid solutions, namely, cases $x = 25$ and $50$\%, are of particular 
interest in terms of practical applications since GaP is scarce in nature and thus expensive, while ZnS is earth-abundant.

\section{Summary}
\label{sec:summary}
We have presented a comprehensive first-principles study of the mixing thermodynamics, and structural, electronic, and photocatalytic 
properties of (GaP)$_{x}$(ZnS)$_{1-x}$ solid--solutions as a function of temperature and composition. Our theoretical approach relies 
on the multi-configurational supercell method, which allows a rigorous statistical treatment of site-occupancy 
disorder as well as calculation of accurate thermodynamic and functional properties of solid solutions. Valuable physical insights
into the atomistic mechanisms behind the thermodynamic and functional features of solid solutions are attained on-the-go 
and interpreted easily in terms of probabilities (e.g., Boltzmann-like occurrence factors). 

We find overall excellent agreement between our calculations and the experimental data reported on the structural and electronic 
properties of GaP--ZnS solid solutions. This good accordance demonstrates the accuracy and reliability of our employed computational 
method. Based on our enery band gap and band alignment results, we predict that compounds $x = 25$, $50$, and $75$\% are very promising 
as photocatalyst materials for generation of hydrogen from water splitting under visible light. Similar theoretical studies to ours may 
be conducted for other encouraging combinations of semiconductor materials (e.g., GaN--ZnO and CdS--ZnS). Our theoretical approach 
is computationally affordable and fully general, hence it has the potential to accelerate the development of optimized materials 
based on semiconductor solid solutions for solar-energy storage; also, to improve the design of materials for a range of other 
applications in which tuning of the opto-electronic behavior is important. We envisage that our approach will promote and assist 
the experimental synthesis of bettered photocatalysts by providing useful insights into their mixing thermodynamics.

\acknowledgments
This research was supported under the Australian Research Council's
Future Fellowship funding scheme (No. FT140100135). Computational resources 
and technical assistance were provided by the Australian Government and the 
Government of Western Australia through the National Computational Infrastructure 
(NCI) and Magnus under the National Computational Merit Allocation Scheme and 
The Pawsey Supercomputing Centre.


\begin{thebibliography}{30}
\bibitem{kudo08} A. Kudo, Y. Miseki,
		 \textit{Chem. Soc. Rev.} \textbf{2009}, \textit{38}, 253.

\bibitem{osterloh08} F. E. Osterloh,
	\textit{Chem. Mater.} \textbf{2008}, \textit{20}, 35. 

\bibitem{tong12} H. Tong, S. Ouyang, Y. Bi, N. Umezawa, M. Oshikiri, J. Ye,
	\textit{Adv. Mater.} \textbf{2012}, \textit{24}, 229.

\bibitem{xing06} C. Xing, Y. Zhang, W. Yan, L. Guo,
	\textit{Int. J. Hydrog. Energy} \textbf{2006}, \textit{31}, 2018.

\bibitem{maeda06} K. Maeda, K. Teramura, D. Lu, T. Takata, N. Saito, Y. Inoue, K. Domen,
	\textit{Nature} \textbf{2006}, \textit{440}, 295. 

\bibitem{yi07} Z. G. Yi, J. H. Ye,
	\textit{Appl. Phys. Lett.} \textbf{2007}, \textit{91}, 254108. 

\bibitem{zou01} Z. Zou, J. Ye, K. Sayama, H. Arakawa,
	\textit{Nature} \textbf{2001}, \textit{414}, 625.

\bibitem{jensen08} L. L. Jensen, J. T. Muckerman, M. D. Newton,  
	\textit{J. Phys. Chem. C} \textbf{2008}, \textit{112}, 3439.

\bibitem{valentin10} C. Di Valentin, 
	\textit{J. Phys. Chem. C} \textbf{2010}, \textit{114}, 7054.

\bibitem{hart13} J. N. Hart, N. L. Allan,  
	\textit{Adv. Mater.} \textbf{2013}, \textit{25}, 2989.

\bibitem{grau14} J. Gonz\'{a}lez-L\'{o}pez, S. E. Ruiz-Hern\'{a}ndez, A. Fern\'{a}ndez-Gonz\'{a}lez, 
	         A. Jim\'{e}nez, N. H. de Leeuw, R. Grau-Crespo,
		 \textit{Geochim. Cosmochim. Acta} \textbf{2014}, \textit{142}, 205. 

\bibitem{dieguez11} O. Di\'eguez, J. ${\rm \acute{I}}$${\rm \tilde{n}}$iguez,
	\textit{Phys. Rev. Lett.} \textbf{2011}, \textit{107}, 057601.

\bibitem{grau17} Y. E. Licea, R. Grau-Crespo, L. A. Palacio, A. C. Faro Jr.,   
	\textit{Catal. Today} \textbf{2017}, \textit{292}, 84.

\bibitem{cazorla18} C. Cazorla, A. K. Sagotra, M. King, D. Errandonea, 
	\textit{J. Phys. Chem. C} \textbf{2018}, \textit{122}, 1267.

\bibitem{bellaiche00} L. Bellaiche, D. Vanderbilt, 
	\textit{Phys. Rev. B} \textbf{2000}, \textit{61}, 7877.

\bibitem{todorov04} I. T. Todorov, N. L. Allan, M. Y. Lavrentiev, C. Freeman, C. E. Mohn,  
                    J. A. Purton,
		    \textit{J. Phys.: Condens. Matt.} \textbf{2004}, \textit{16}, S2751.

\bibitem{allan01} N. L. Allan, G. D. Barrera, R. M. Fracchia, M. Yu. Lavrentiev, M. B. Taylor, 
	          I. T. Todorov, J. A. Purton,  
	          \textit{Phys. Rev. B} \textbf{2001}, \textit{63}, 094203.

\bibitem{wei90} S.-H. Wei, L. G. Ferreira, J. E. Bernard, A. Zunger,
	\textit{Phys. Rev. B} \textbf{1990}, \textit{42}, 9622.

\bibitem{grau07} R. Grau-Crespo, S. Hamad, C. R. A. Catlow, N. H. de Leeuw,
	\textit{J. Phys. Condens. Matter} \textbf{2007}, \textit{19}, 256201. 

\bibitem{grau12} R. Grau-Crespo, U. V. Waghmare,
                 \textit{Molecular Modeling for the Design of Novel Performance Chemicals and
                 Materials} \textbf{2012}, ISBN:1439840784, CRC Press, pp.~299-322.

\bibitem{shintani73} A. Shintani, S. Minagawa,
	\textit{J. Phys. Chem. Solids} \textbf{1973}, \textit{34}, 911.
 
\bibitem{sonomura73} H. Sonomura, T. Uragaki, T. Miyauchi,
	\textit{Jpn. J. Appl. Phys.} \textbf{1973}, \textit{12}, 968.

\bibitem{park14} K. Park, J. A. Lee, H. S. Im, C. S. Jung, H. S. Kim, J. Park, C.-H. Lee,  
	\textit{Nano Lett.} \textbf{2014}, \textit{14}, 5912.

\bibitem{yeh92} C.-H. Yeh, Z. W. Lu, S. Froyen, A. Zunger, 
	\textit{Phys. Rev. B} \textbf{1992}, \textit{46}, 10086.

\bibitem{pbe96} J. P. Perdew, K. Burke, M. Ernzerhof,
	\textit{Phys. Rev. Lett.} \textbf{1996}, \textit{77}, 3865.

\bibitem{vasp} G. Kresse, J. F\"urthmuller,
	\textit{Phys. Rev. B} \textbf{1996}, \textit{54}, 11169.

\bibitem{bloch94} P. E. Bl\"ochl,
	\textit{Phys. Rev. B} \textbf{1994}, \textit{50}, 17953.

\bibitem{hse06} A. V. Krukau, O. A. Vydrov, A. F. Izmaylov, G. E. Scuseria,
	\textit{J. Chem. Phys.} \textbf{2006}, \textit{125}, 224106.

\bibitem{kresse95} G. Kresse, J. Furthm\"uller, J. Hafner,
	\textit{Europhys. Lett.} \textbf{1995}, \textit{32}, 729.

\bibitem{alfe09} D. Alf\`e,
	\textit{Comp. Phys. Commun.} \textbf{2009}, \textit{180}, 2622.

\bibitem{crystal09} R. Dovesi, R. Orlando, B. Civalleri, C. Roetti, V. R. Saunders, C. M. Zicovich-Wilson,
	\textit{Z. Kristallogr.} \textbf{2005}, \textit{220}, 571. 

\bibitem{becke93} A. D. Becke, 
	\textit{J. Chem. Phys.} \textbf{1993}, \textit{98}, 5648.

\bibitem{walle02} A. van de Walle, G. Ceder, 
	\textit{Rev. Mod. Phys.} \textbf{2002}, \textit{74}, 11.

\bibitem{cazorla17} C. Cazorla, O. Di\'eguez, J. ${\rm \acute{I}}$${\rm \tilde{n}}$iguez, 
	\textit{Sci. Adv.} \textbf{2017}, \textit{3}, e1700288.

\bibitem{cazorla13} C. Cazorla, J. ${\rm \acute{I}}$${\rm \tilde{n}}$iguez,
	\textit{Phys. Rev. B} \textbf{2013}, \textit{88}, 214430.

\bibitem{cazorla17b} C. Cazorla, J. Boronat, 
	\textit{Rev. Mod. Phys.} \textbf{2017}, \textit{89}, 035003.

\bibitem{becker00} U. Becker, A. Fernandez-Gonzalez, M. Prieto, R. Harrison, A. Putnis, 
	\textit{Phys. Chem. Miner.} \textbf{2000}, \textit{27}, 291.

\bibitem{jones73} H. Jones, 
	\textit{Rep. Prog. Phys.} \textbf{1973}, \textit{36}, 1425.  

\bibitem{heyd05} J. Heyd, J. E. Peralta, G. E. Scuseria, R. L. Martin, 
	\textit{J. Chem. Phys.} \textbf{2005}, \textit{123}, 174101.

\bibitem{moses11} P. G. Moses, M.  Miao, Q. Yan, C. G. Van de Walle,  
	\textit{J. Chem. Phys.} \textbf{2011}, \textit{134}, 084703.

\bibitem{hamad02} S. Hamad, S. Cristol, C. R. A. Catlow, 
	          \textit{J. Phys. Chem. B} \textbf{2002}, \textit{106}, 11002.

\bibitem{wright98} K. Wright, G. W. Watson, S. C. Parker, D. J. Vaughan, 
	           \textit{Am. Mineral.} \textbf{1998}, \textit{83}, 141.

\bibitem{hayashi82} K. Hayashi, M. Ashizuka, R. C. Bradt, H. Hirano, 
	            \textit{Mater. Lett.} \textbf{1982}, \textit{1}, 116.

\bibitem{vandewalle03} C. G. Van de Walle, J. Neugebauer, 
	\textit{Nature} \textbf{2003}, \textit{423}, 626. 
\end{thebibliography}
\end{document}